\documentclass[apjl]{emulateapj}

\def\pdeg{\ifmmode $\setbox0=\hbox{$^{\circ}$}\rlap{\hskip.11\wd0 .}$^{\circ}
          \else \setbox0=\hbox{$^{\circ}$}\rlap{\hskip.11\wd0 .}$^{\circ}$\fi}
\def\arcs{\ifmmode {^{\scriptscriptstyle\prime\prime}}
          \else $^{\scriptscriptstyle\prime\prime}$\fi}
\def\arcm{\ifmmode {^{\scriptscriptstyle\prime}}
          \else $^{\scriptscriptstyle\prime}$\fi}
\newdimen\sa  \newdimen\sb
\def\parcs{\sa=.07em \sb=.03em
     \ifmmode \hbox{\rlap{.}}^{\scriptscriptstyle\prime\kern -\sb\prime}\hbox{\kern -\sa}
     \else \rlap{.}$^{\scriptscriptstyle\prime\kern -\sb\prime}$\kern -\sa\fi}
\def\parcm{\sa=.08em \sb=.03em
     \ifmmode \hbox{\rlap{.}\kern\sa}^{\scriptscriptstyle\prime}\hbox{\kern-\sb}
     \else \rlap{.}\kern\sa$^{\scriptscriptstyle\prime}$\kern-\sb\fi}
\def\mo{\ifmmode ^{-1}\else $^{-1}$\fi}

\begin{document}

\title{Spitzer observations of the dusty warped disk of Centaurus\,A}

\author{Alice C. Quillen}
\affil{Department of Physics and Astronomy,
University of Rochester, Rochester, NY 14627}
\email{aquillen@pas.rochester.edu}
\author{Mairi H. Brookes, Jocelyn Keene, Daniel Stern,
Charles R. Lawrence, Michael W. Werner}      
\email{Mairi.H.Brookes@jpl.nasa.gov}
\email{jkeene@caltech.edu} 
\email{charles.lawrence@jpl.nasa.gov}
\email{Michael.W.Werner@jpl.nasa.gov}
\email{stern@zwolfkinder.jpl.nasa.gov}
\affil{Jet Propulsion Laboratory, 4800 Oak Grove Drive, Pasadena, CA 91109}

\begin{abstract}

{\it Spitzer} mid-infrared images of the dusty warped disk in the galaxy
Centaurus\,A show a parallelogram-shaped structure.  We successfully model the
observed mid-infrared morphology by integrating the light from an emitting,
thin, and warped disk,  similar to that inferred from previous kinematic
studies.   The models with the best match to the morphology lack dust emission
within the inner 0.1 to 0.8\,kpc, suggesting that energetic processes near the
nucleus have disturbed the inner molecular disk, creating a gap in the
molecular gas distribution.

\end{abstract}


\section{Introduction}

Centaurus\,A (NGC\,5128) is the nearest of all the giant radio galaxies.  
Because of the disk of gas and
dust in its central regions,  Centaurus\,A is suspected to be the product of a
merger of a small gas rich  spiral galaxy with a larger elliptical galaxy
\citep{baade}.   Numerical simulations of such mergers  predict large 
shell-like features \citep{hernquist88} that
have been observed in Centaurus\,A over a large range of radii
\citep{malin83,peng02}.  Some contain atomic and molecular gas, implying that
they were formed from material stripped from a galaxy relatively recently, fewer
than a few galactic rotation periods ago \citep{schiminovich94,charmandaris}.
Tidal features associated with this debris have also been identified,
supporting the relatively short  (few times $10^8$ years) estimated timescale
since the merger took place  \citep{peng02}.

In its central regions, NGC\,5128 exhibits a well recognized, optically-dark
band of absorption across its nucleus. This dusty disk was first modeled as a
transient  warped structure by \citet{tubbs80}.  
\citet{bland86}, \citet{bland87}, \citet{quillenCO}, and \citet{nicholson92}
found that the kinematics of the ionized and molecular gas are well modeled by
a warped disk composed of a series of inclined  connected rings undergoing
circular motion  (as also explored for other galaxies with peculiar morphology 
by \citealt{steiman92}). The model explored by Quillen, Graham \& Frogel (1993) 
modified the kinematic model by \citet{quillenCO} to fit the morphology of the
absorptive, dusty disk seen in near-infrared images  and proposed a timescale
of about $200$ million  years since the core of an infalling spiral galaxy
reached and merged with the elliptical galaxy nucleus. An initially flat disk,
misaligned with the galaxy principal symmetry axis, becomes increasingly
corrugated as a function of time. The short timescale estimated since the
merger in NGC\,5128 is approximately consistent with the timescale suggested
by the presence of tidal debris and by the shell-like features containing atomic
hydrogen \citep{schiminovich94,peng02} and molecular gas \citep{charmandaris}.  
An alternative model accounting for the warped disk is the polar ring model by
\citet{sparke96},  consistent with the polar orbit of the disk implied by the
galaxy isophotes in the outer galaxy \citep{malin83,haynes83,peng02} but
requiring a longer timescale to  account for the twist of the warp.

More recent observations of the central region include submillimeter imaging
with the Submillimeter Common User Bolometer Array (SCUBA) and mid-IR imaging
with the Infrared Camera (ISOCAM) on the {\it Infrared Space Observatory (ISO)} 
satellite \citep{mirabel99,leeuw02}.   At these wavelengths, the dusty disk is
seen in  emission rather than absorption. At 100--200\,pc from the nucleus the
galaxy contains a molecular circumnuclear  disk that has been studied in
molecular line emission \citep{israel90,israel91} and resolved in Pa\,$\alpha$
emission \citep{schreier96,marconi01}.  For a recent summary of the wealth of
observational studies carried out on  this peculiar and active nearby galaxy,
see the comprehensive review by \citet{israel}. Based on the discussion by
\citet{israel},  we adopt a distance to Centaurus\,A of $3.4$\,Mpc.  At this
distance $1\arcm$ on the sky corresponds to $\sim 1$\,kpc.

In this manuscript we focus on the geometry of the dusty disk in Centaurus\,A as
seen from {\it Spitzer\/} Infrared Array Camera (IRAC) images.  These images
resolve the structure of the disk more clearly than previous observations 
and also show the disk out to larger radii.  They provide us with an
opportunity to study the geometry of the warped disk in much higher detail than
previously possible. Observations are described in \S\,2. Our geometric
model is described in \S\,3.  A discussion and summary follow in \S\,4.


\section{Observations}

\begin{figure*}
\plotone{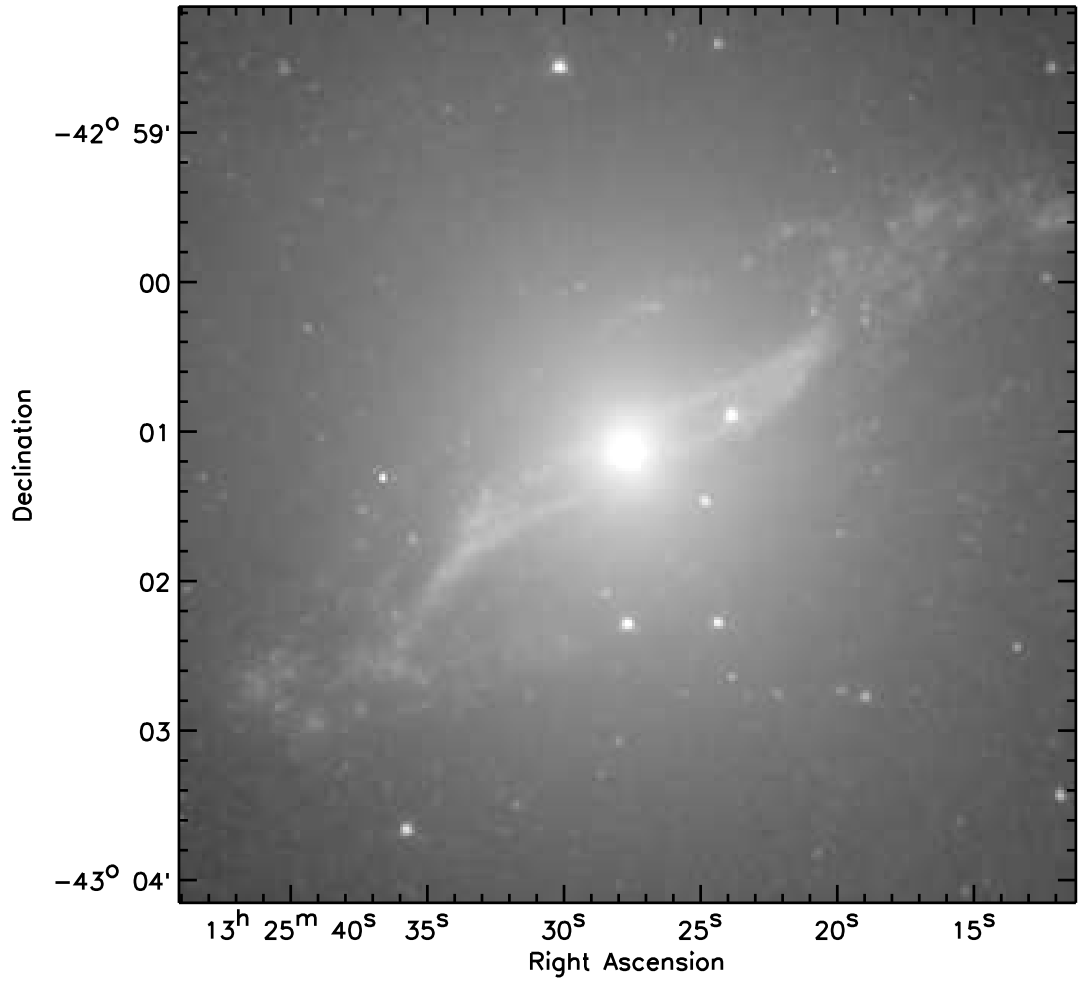}
\figcaption{
a) The central region of the image at 3.6\,$\mu$m taken
by the {IRAC} camera on board the {Spitzer Space Telescope}.
The emission is shown on a log scale.
At 3.6 and 4.5\,$\mu$m starlight peaking near the galaxy center 
is seen in addition to emission from dust.
At longer wavelengths emission from dust is primarily seen.
\label{fig:iracband1}
}
\end{figure*}

\begin{figure*}
\plotone{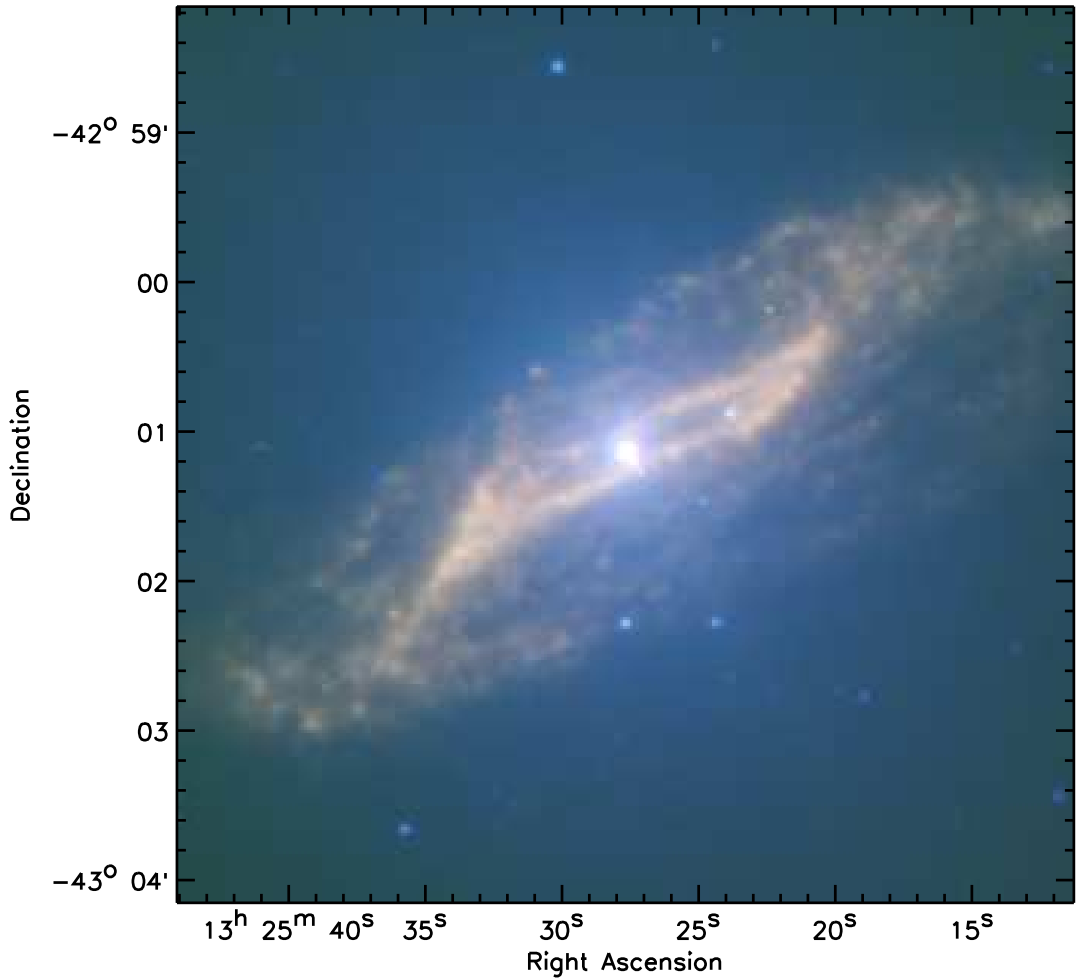}
\figcaption{
The same region as Fig.~\ref{fig:iracband1} 
but a color composite of the IRAC images at 3.6, 5.8 and $8.0\mu$m.
This figure is shown on a log scale.
The black and white version of this figure only shows the emission at 
at 8.0\,$\mu$m. 
\label{fig:iracband4}
}
\end{figure*}

Figures~\ref{fig:iracband1} and \ref{fig:iracband4} 
present images of NGC\,5128 taken on 2004
February~10 in the 3.6, 4.5, 5.8 and 8.0\,$\mu$m broadband filters (channels
1-4) of the {\it  Spitzer\/} Infrared Array Camera ({IRAC}; \citealt{fazio}).
In each filter, {\it fixed cluster\/} observing mode was used for these
observations to produce a map of $5\times6$ points, at which five dithered
exposures were taken. The exposure time per frame was 12\,s.  Additional shorter
exposure, 0.4\,s frames, were also taken to correct for possible saturation in
the longer exposure frames. 
The coverage of the map at each position on the sky varies between three and
ten frames, with an average coverage of six frames, corresponding to an
exposure time of 72\,s.

Before mosaicing, the {\it basic calibrated data\/} (BCD) frames were corrected
for artifacts using the IRAC artifact mitigation code (excluding the pulldown
correction) available from the {\it Spitzer\/} Science Center contributed
software
pages\footnote{http://ssc.spitzer.caltech.edu/archanaly/contributed/browse.html}.
The final mosaiced images were produced from the dithered frames by applying
the {\it MOPEX} software\footnote{http://ssc.spitzer.caltech.edu/postbcd/} to
the corrected BCD frames.

Here we focus on the central $5\arcm\times 5\arcm$ region in the final mosaics. 
The mosaiced images
provide a plate scale with 1\parcs2 pixels; FWHMs of the point spread functions
are  1.7, 1.7, 1.9, 2\parcs0 in channels 1--4, respectively. 
The rms noise levels in these images are 0.025, 0.024, 0.070, and
0.060\,MJy\,sr$^{-1}$ in channels 1--4, respectively.  The above
sensitivities agree with the predictions of the {\it SENS-PET\/} sensitivity
estimator in longer wavelengths channels.  The 3.6\,$\mu$m channel contains
emission from the stellar component of the galaxy that extends over a large
portion of the field of view. This makes the 3.6\,$\mu$m image about half as
sensitive as the {\it SENS-PET\/} prediction for the dust emission. 


The {IRAC} images presented here are deeper than previous ISOCAM images
(\citet{mirabel99}), and have higher angular resolution than
previous submillimeter images (\citep{mirabel99,leeuw02}).  In the inner few
arcminutes, the {IRAC} images exhibit a parallelogram shape in emission   
(see Fig.~\ref{fig:iracband4}).  This shape, previously seen at lower angular
resolution,  was interpreted as an S-shape, possibly associated with shocks 
\citep{mirabel99,leeuw02}.  A parallelogram morphology has been seen previously 
in other galaxies.  For example, dust absorption features in the SO galaxy
NGC\,4753 exhibit a parallelogram shape, and have been modeled with a warped
twisted disk by \citet{steiman92}.  When an optically thin warped disk is seen
in emission, the edges of folds in the disk correspond to regions of higher
surface brightness. Multiple folds are seen along the line of sight  (e.g.,
\citealt{bland86,bland87,quillen,nicholson92})  implying that some parts of the
disk are seen through other parts of the disk. In this case the morphology
would not be nearly  symmetric across the origin ($r \to -r$), as observed, 
unless the outer disk was nearly optically thin in the mid-infrared. We infer
that the disk probably has a low normal optical depth in the mid-infrared (if
observed face-on), though the bright edges of the folds  and individual clumps
in the disk may not be optically thin.  The parallelogram is present in all four
IRAC bands, though it is most prominent compared to the diffuse emission from
starlight in the longer wavelength $5.8$ and $8\mu$m bands.

\begin{figure*}
\plotone{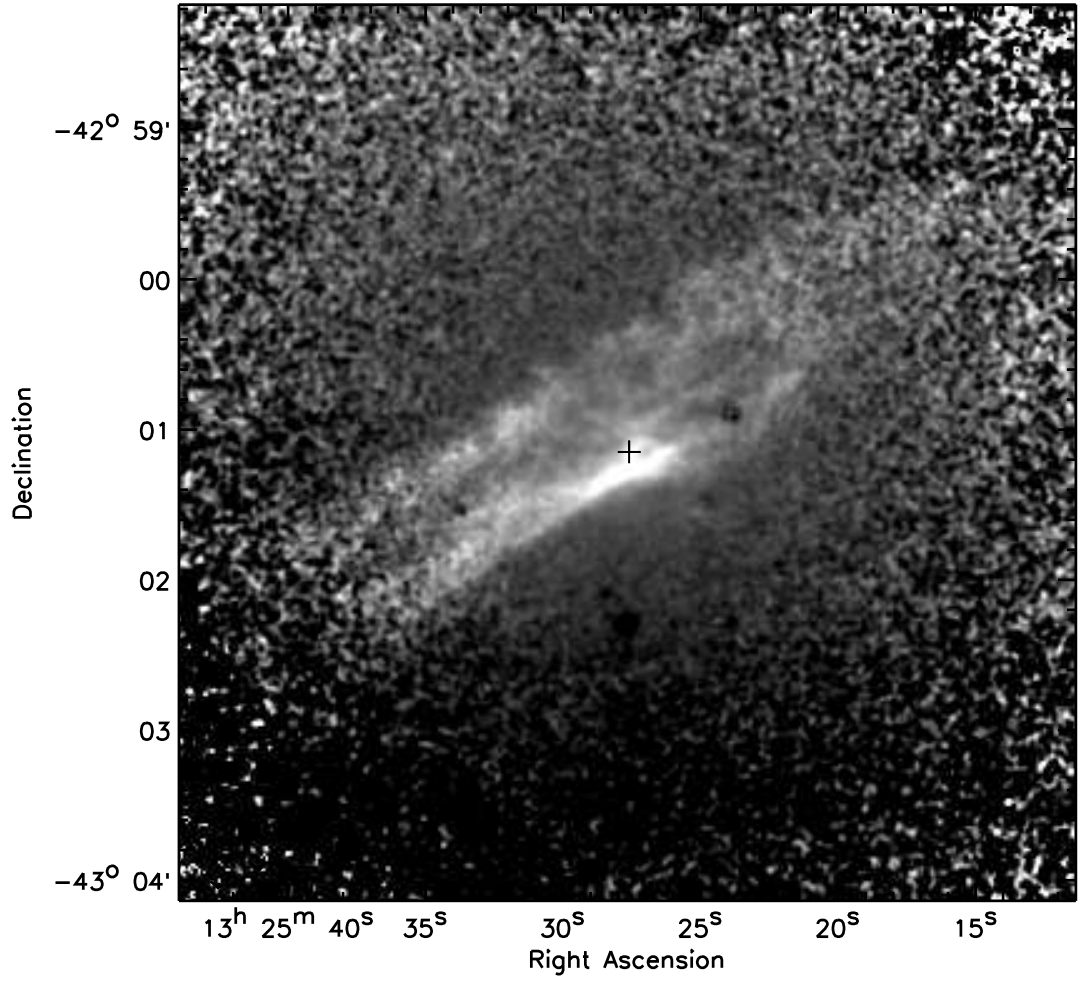}
\figcaption{
Color map showing the log of the H-band divided by the J-band 
{2MASS} images by \citet{jarrett}.
Lighter shadding corresponds to regions of heavier extinction.
Folds in the disk both nearer and more distant the observer
are seen in the mid-infrared images. However folds nearer
the observer absorb more background starlight and so
are more prominent in the near-infrared images. 
\label{fig:colors}
}
\end{figure*}

In visible or near-infrared bands, folds can correspond to regions where the
absorption from dust obscures background starlight.  In this case folds that
are closer to the observer absorb more background starlight from the galaxy.   
Thus some of the features seen in emission in the {IRAC} images resemble and
coincide with absorption bands previously seen in near-infrared images.  
In Figure~\ref{fig:colors} we show a color map 
made from 2\,Micron All Sky Survey
(2MASS) images from the 2MASS large galaxy atlas \citep{jarrett}.  The
south-eastern edge of the parallelogram in the {IRAC} images  lies in the same
location as the absorption band $\sim 10\arcs$ to the south east of the nucleus
prominent in the near-infrared images.  This is consistent with the study by
\citet{leeuw02}, who compared submillimeter images to near-infrared images. The
edge of emission $\sim 1\arcm$ to the north of the nucleus (at a surface
brightness below that of the parallelogram) in the   8.0\,$\mu$m {IRAC} image
corresponds to the top of the dust lane  prominent in optical images of the
galaxy. This corresponds to the extinction  feature in the near-infrared color
map to the north of the nucleus. From comparing the IRAC images to the 2MASS
images we infer that the southern side of the parallelogram is closer to the
observer than the northern side.  Because the dust on this side lies in front
of the plane perpendicular to the line of sight containing the galaxy nucleus,
it absorbs more background starlight and so causes a deeper band of extinction
in the near-infrared images.  The northern side of the parallelogram lies on the
opposite side and so is not seen in the near-infrared images.  Likewise the
oval edge of emission $\sim 1\arcm$ to the north  of the galaxy nucleus  (see
Fig.~\ref{fig:iracband4})  is nearer the observer than that on the opposite
edge $\sim 1\arcm$ to the south of the nucleus, and only the northern side
causes an absorption feature in the near-infrared images and color maps  (see
Fig.~\ref{fig:colors}).



\section{Modeling of the warped disk}

At many observed wavelengths, the morphology of Centaurus\,A is well reproduced
by geometric models of a warped disk \citep{bland86,bland87, quillenCO,
nicholson92, quillen, sparke96}.   We describe and extend such modeling here.

A warped disk can be described as a series of tilted rings, each with a
different radius $r$.   We assume that the gas and dust are evenly distributed
on each ring and each ring is smoothly connected to those at larger and smaller
radii. We follow the notation and framework used by \citet{quillenCO,quillen} 
to describe the geometry of a warped disk.   Each ring is described by two
angles, a precession angle $\alpha(r)$ and an inclination angle $\omega(r)$. 
These angles are given with respect to an assumed principal axis of the
underlying elliptical galaxy.  This axis requires two angles to describe,
$\chi$, corresponding  to the position angle counter clockwise from north of the
axis on the sky,  and an inclination angle, $\vartheta$, that describes the
tilt of this  galactic axis with respect to the line of sight (see Fig.~6 by
\citealt{quillen}).  We assume that the galaxy is axisymmetric and not triaxial
so a third angle is not required to describe its orientation.  We also assume
that the galaxy shape is fixed, and not tumbling.

Our description for the ring projection angles can be related to those of
previous works. While \citet{quillenCO,quillen} described the angles of the
warped disk  with respect to the principal axis of the underlying  elliptical
galaxy, \citet{bland87} and \citet{nicholson92} matched the H\,$\alpha$
velocity field with a tilted ring  model describing the orientation of the
rings with respect to the line of sight.  This model fit the velocity field by
adjusting the ring inclination  as a function of radius, $i(r)$, and position
angle counter clockwise from north, $p(r)$.


To produce a model image of the mid-infrared emission, we must consider all
emitting and absorbing regions along the line  of sight at each position on the
sky. This is much simpler when the system is optically thin.  In  this case, we
sum all emission along the line of sight at each position on the sky.  The near
symmetry of the disk suggests that we can use an optically thin approximation
to model this disk.  While individual clouds could contain optically thick
regions, the bulk of the emitted mid-infrared light is likely to reach the
observer. When the disk is optically thin, brighter areas correspond to regions
that appear folded from the perspective of the viewer.  Here we neglect
emission from  a spherical component associated with the stars and only consider
emission from a thin but warped disk.

Our numerical procedure begins by randomly sampling $x,y$ positions in the
plane perpendicular to the principal axis of the elliptical galaxy.  At each
position we compute a $z$-coordinate based on a smooth (spline) function for
the disk precession and inclination angles, $\alpha(r)$ and $\omega(r)$.   The
points that specify these spline functions are listed in Table 1. The
coordinates for each point are then rotated into the viewer's frame.  Points
along each line of sight are summed to produce a model image.  For the functions
$\alpha(r)$ and $\omega(r)$, we began with those from
\citet{quillen} and slightly varied the angles at different radii to achieve a
better match to the observed morphology.  Matching was done by visual
comparison to the {IRAC} images.  No minimizing fit to the image data values
was done. Our procedure is adequate to understand the projection affects
associated with an emitting warped disk.  Future and more intensive modeling
would be required to do a multi-dimensional  and multi-wavelength fit.

To construct a model, we must consider the thickness of the disk and its
brightness distribution.  Because the disk is not infinitely thin we randomly
chose slight offsets from the $z$-coordinate computed from the precession and
inclination angles.   The size of the offset depends on a disk aspect ratio
$k(r) = h/r$, where $h(r)$ is the disk thickness as a function of radius.  We
assumed for the disk aspect ratio a power law form,
$k(r) = k_{50}(r/50\arcs)^{\beta_z}$.  The intensity of the emission
contributed at each point was then computed assuming that the disk volume
emissivity integrated through the disk vertically is a power law function of
radius, $\epsilon(r) \propto r^{-\beta_e}$. We also allowed an inner gap in the
radial surface brightness profile, $r_{gap}$. The contribution from the active
nucleus and the $\sim 100$\,pc  circumnuclear disk are not taken into account in
our model so by a gap  we mean a deficit within $r_{gap}$ and that of the
circumnuclear disk at $\sim 6\arcs$. 

A model intensity image computed as described above  is shown in
Figure~\ref{fig:model} along with the $8.0\mu$m {IRAC} image of the galaxy. The
numerical parameters of this model are summarized  in Table \ref{mod_params}.
The angle precession and inclination angular functions, 
$\alpha(r)$ and $i(r)$, are shown in Figure~\ref{fig:alpha} as a solid and
dashed lines, respectively. The precession angle used previously by
\citet{quillen} from matching the near-infrared morphology is shown as a dotted
line in this figure and is similar to our best matching model.  The {\it
Spitzer\/} data allow us to better study the outer parts of the disk than
possible with the old $^{12}$CO(2-1) spectra and near-infrared images. Thus
there are some differences in the precession angle $\alpha(r)$ between our old
model and our newer one, as shown in Figure~\ref{fig:alpha}. To better match the
morphology in  the outer parts of the disk we allowed the disk to twist further
(to higher a higher value of $\alpha$) before decreasing at larger radius.

Our model predicts that the disk alternates between having the southern side
and northern side nearest, with folds both above  the nucleus.  The
near-infrared images show a strong dust absorption feature to the southwest of
the nucleus corresponding to  an inner fold, and a weaker feature north of the
nucleus corresponding to an outer fold \citep{quillen}.  In
Figure~\ref{fig:ring} we show the rings comprising our warp model,  projected
on the sky. The nearer semi-circle of each ring is shown in red, whereas the
more distant semi-circle is shown in blue.  The disk at $r \sim 60\arcs$ is nearest
the observer on the south-east side.  This region corresponds to a fold in the
disk that is  seen in the near-infrared color map, $10\arcs$ to the south-east of
the nucleus (see Fig.~\ref{fig:colors}).  At $r \gtrsim 100\arcs$ the northern
side of the disk is again nearest the observer.   This corresponds to the
northern dust lane seen in the the optical images and the near-infrared color
map about $45\arcs$ north of the nucleus.


\subsection{Comparison to previous studies on the geometry of the warped disk}

In Figure~\ref{fig:nichol}a,b,  we show the inclination and position angles  as
a function of radius with respect to the viewer; functions that can be directly
compared to those found by \citet{nicholson92} from their kinematic fits to the
H\,$\alpha$  velocity field (see their Fig.~9a,b).  \citet{nicholson92} adopted
a distance of 3\,Mpc to the galaxy, so their linear scale is 15\% different than
ours; we have corrected for this difference by rescaling the given
distances.   We see from Figure~\ref{fig:nichol} that the position angles of
their tilted ring fit are quite similar to those predicted by our model.  

The warp shape originally designed to match the velocity field in molecular gas
\citep{quillenCO} resembled that which matched the H\,$\alpha$ velocity field
by \citet{nicholson92}. Both models 
included an inclined disk that tilted so that gas
rings at different radii alternated between being retrograde and prograde with
respect to the observer.   This was also a characteristic of the high
inclination warp model for NGC\,4753 by \citet{steiman92}.   The tilted ring fit
by \citet{nicholson92} to the velocity field did not specify   which side of
the each ring was closest the viewer.  However
\citet{bland86}, \citet{bland87}, \citet{nicholson92}, and \citet{quillen} used
the visible and near-infrared images to break this degeneracy. 
The disk is tilted so that it is nearer
the viewer on the southern side for intermediate radii;  accounting for the
southern edge of the dust lane seen in optical images, and nearer the viewer on
the northern side at largest radii to account for the northern dust lane seen 
in the optical and near-infrared images.   These flips in orientation can be
seen from Figure~\ref{fig:nichol}b. They occur where the inclination with
respect to the viewer crosses $90^\circ$ at which point the disk is edge-on.

The model by \citet{quillen} had a somewhat lower inclination for the galaxy
principal axis, $\vartheta = 65^\circ$, instead of $75^\circ$, used here. Their
model also had a higher disk inclination with respect to the galaxy principal
axis at smaller radii.  There is redundancy in the  model between the
inclination of the disk with respect to the galaxy axis,   $\omega(r)$,  and
the tilt of the galaxy principal axis, $\vartheta$.  The maximum and minimum
ring inclinations correspond to values set by $\vartheta \pm  \omega$ (see
discussion by \citealt{quillenCO}).  To exhibit the parallelogram shape, the
disk inclination with  respect to the viewer must go above and below $90^\circ$
causing folds in the disk.   This restricts the models to  a range of values
$\vartheta + \omega \sim 100^\circ$.

Here, we find a somewhat better match  (see Fig.~\ref{fig:bigmodel}) to the
mid-infrared images  with a model that has decreasing inclination (with respect
to our estimated galaxy principal axis) at larger radii.  However this decrease
may be spurious.  To compute the model at sufficient resolution to compare to
the inner region, we cannot well sample points over a large region.  By
accounting for every observed feature we may have  achieved a better morphology
match with a model that has exaggerated variations in the ring angles.  In
other words, the angular variations may have been compressed into a smaller 
radii than they should be.

To be more certain of the orientation of the outer disk,  better kinematic
constraints on this outer disk are needed.  By specifying a choice for the
principal axis of the galaxy, we have chosen to describe the geometry of the
warped disk with respect to a particular axis.   This choice helped
\citet{quillen} compare the shape of the disk to the predictions of simple
merger and precessing ring models. However, it is not necessarily significant
that the disk inclination as measured with respect to our assumed galaxy axis
varies slowly with radius. An additional complexity not considered with our
choice of projection angles is that the galaxy may be triaxial, or
significantly vary in shape and orientation with radius (e.g., as considered by
\citealt{arnaboldi94}).   The galaxy could be in the process of dynamic
relaxation following the merger. Analytical models or one-dimensional
integrations fail to capture the complexity of more detailed numerical
simulations, particular for near equal mass mergers (e.g.,
\citealt{mihos99,mihos_hernquist96}). Imaging and kinematic studies of the
outer galaxy suggest that these additional degrees of freedom are important 
\citep{malin83,schiminovich94,hui95,peng02}.

\subsection{In context with the dynamical warp models} 

The model precession angle $\alpha(r)$ reaches a maximum at $r\sim 100\arcs$,
decreasing at both larger and smaller radii (see Fig.~\ref{fig:alpha}).   This
implies that the corrugated disk is twisted in one direction for $r\lesssim
100\arcs$ and in the opposite direction for $r\gtrsim 100\arcs$. Since the
near-infrared morphology depends on which side of each ring is closer to the
observer, and our model is similar to that used to match the near-infrared
images  (also see discussion by \citealt{bland87,nicholson92}), this peculiar
change in the handedness of the twist is likely to be real. When the handedness
of the  twist is the same at all radii, a nested set of parallelogram shapes
can be seen (e.g., as in M84; \citealt{quillenm84}).   However in Centaurus\,A, 
an outer oval is seen in the outer disk (see Fig.
\ref{fig:bigmodel} and \ref{fig:ring}) that is slanted in the opposite
direction as the inner parallelogram. This difference in the direction of the
slant is a feature of  the sign change in the slope of the precession angle.

The precession rate of a gas ring inclined with respect to the underlying galaxy
undergoing circular motion  is proportional to the ellipticity of the galactic
gravitational potential  and the angular rotation rate of the ring  (e.g.,
\citealt{tubbs80,sparke84}). If the self gravity of the ring is important  then
it too can affect the precession rate \citep{sparke86,arnaboldi94,sparke96}. The
direction of precession depends on  the orientation of the ring and galaxy
(whether a polar ring or not) and whether  the galaxy is prolate or oblate.  
Most models assume that following a merger the gas and dust  are distributed in
a plane that is misaligned with the principal axes of the galaxy. To be
consistent with the direction of the twist in the outer parts of Centaurus\,A 
(decreasing $\alpha(r)$ at $r\gtrsim 100\arcs$), the galaxy can either be
prolate  and the ring located near the equatorial plane \citep{quillen} or the
galaxy can be oblate and the ring would be nearly polar
\citep{sparke96}.


The reversal in the sign of the slope ($d\alpha/dr$) is unlikely to be caused
solely by the shape of the rotation curve.  
The predicted rotation curve of Centaurus\,A increases at smaller radius; all
the way to a radius of $10\arcs$ (see Fig.~13 by \citealt{marconi06}).  
Consequently the angular rotation rate should increase as the radius decreases
nearly all the way to the galaxy center.  The rotation curves used by
\citet{quillen} and \citet{sparke96} were nearly solid-body at small radii, and
so underestimated the angular precession rate at small radii (within $\sim
1.5\arcm$ of the nucleus).

The reversal in the sign of the slope ($d\alpha/dr$)  could be due to a drop in
the galaxy eccentricity (exploited by both \citealt{quillen,sparke96}) that
would reduce the precession rate at small radii.    To account for the change
in slope of $\alpha$ at $r\sim 100\arcs$, \citet{quillen} assumed a sharp
cutoff  in the ellipticity of the gravitational potential of the galaxy as a
function of radius. For $r<80\arcs$, the potential ellipticity was much reduced
in the model by \citet{quillen}  compared to that outside it.   The galaxy
isophotes are best viewed in the short  wavelength $3.6\mu$m {IRAC} image where
the stellar light most contributes to the flux and the extinction and emission
from dust is minimized compared to that at shorter wavelengths in the
near-infrared. Galaxy isophotal contours for this image are shown in
Figure~\ref{fig:ch1}.

We first examine the isophotes at $3.6\mu$m to search for evidence of the large
scale stellar bar proposed by \citet{mirabel99}. If such a bar were responsible
for the parallelogram shape in the mid-infrared images, the bar would be
viewed at intermediate inclinations (not edge-on), and so should be
evident in the isophote shapes. However, the isophotes do not exhibit a region
of flat surface brightness or a change in ellipticity or position angle over a
short range in radius.  These features would be expected at the end of a
stellar bar.    We conclude that there is no evidence for a large scale, few
kpc sized,  stellar bar at the heart of Centaurus\,A.

The galaxy isophotal ellipticity does decrease in the inner regions.  By fitting
contours to the $3.6\,\mu$m image outside the emitting disk, we measured a
galaxy isophotal ellipticity for the stellar distribution ($\epsilon = 1 - b/a$,
where $b$ and $a$ are the semi-minor and semi-major axes) of $\epsilon \sim 0.1$
at $r\sim 4\arcm$, $\epsilon \sim 0.05$ at $r \sim 2\arcm$, and $\epsilon \sim
0$ at $r<1\arcm$.  Because the gravitational potential is a convolution of the
density profile with a $1/r$ function, the gravitational potential contours are
smoother than the associated density distribution.  However, the isophotes are
likely to be rounder and smoother than the actual density distribution. While
the drop in background galaxy ellipticity exploited  by \citet{quillen,sparke96}
in their dynamical models is real, it may not be sharp enough  to account for
the abrupt drop in the precession rate inferred in the central region of the
galaxy from the change in the slope of the precession angle $\alpha$ at $r \sim
100\arcs$.

We now consider the role of the mass in the disk.  \citet{sparke96} showed that
the self gravity of the disk could increase the precession rate and vary the
precession axis in the central region.  \citet{sparke96} only considered the
mass in the molecular and atomic gas  components.  \citet{quillen} noted that
there were extensions in the K-band isophotes that were not reproduced by the
purely absorptive disk model.  The K-band isophotes are extended at a radius of
about $60\arcs$ in the 15 mag/arcsec$^2$ contour (also see the dereddened
contour map shown in Figure~9 by \citealt{marconi06}).  We can compare the
mass in a possible disk gas and stellar component to that in the underlying
galaxy.  With a circular rotational velocity of $250$\,km\,s\mo, a total mass
of $\sim 10^{10}\,M_\odot$ is enclosed within a radius of 1\,kpc.  The mass in
molecular gas in the same region is a few times $10^8\,M_\odot$ (Phillips et
al.~1987, corrected for the difference in the assumed distance). The level of
15\,mag\,arcsec$^{-2}$ in K-band corresponds to a  surface density of
$4500\,M_\odot\,$pc$^{-2}$ assuming a mass to light ratio at K-band of 0.5 
(e.g., \citealt{silge05}).  
The actual surface density would be $\sim 10$ times lower than this, taking
into account the mean disk axis ratio on the sky.  This surface density can be
compared to the estimated surface density of molecular gas, or a few hundred
$M_\odot$\,pc$^{-2}$. This suggests that at least an equal mass exists in stars
in the disk as in gas.  

Comparing the total mass in gas and stars in the disk to that in the underlying
spherical component, we estimate that a few percent of the total mass within
1\,kpc lies in the disk.  The mass in the disk is likely to be a few times
larger than that used by \citet{sparke96} to account for the disk geometry.  
We support the finding by \citet{sparke96} that the self-gravity in the disk is
important, and so should significantly affect the disk precession rate. 
Future modifications of the prolate model should take this into account, as the
self-gravity in the disk could change the direction of precession in this
region. 
If the outer disk is prolate, then a reversal in the twist could be due to the
self-gravity of the disk in the inner region.  If the outer galaxy is oblate,
then the ring is polar and the model by \citet{sparke96} would account for the
reversal in the twist of the disk.
Both dynamical models could be updated to include better estimates of the mass
in the disk, the galaxy isophotes at 2.2 and $3.6\,\mu$m, and an improved
rotation curve based on the light distribution.

\subsection{A gap in the dusty disk between a radius of $6\arcs$ and $50\arcs$}

\begin{figure*}
\plotone{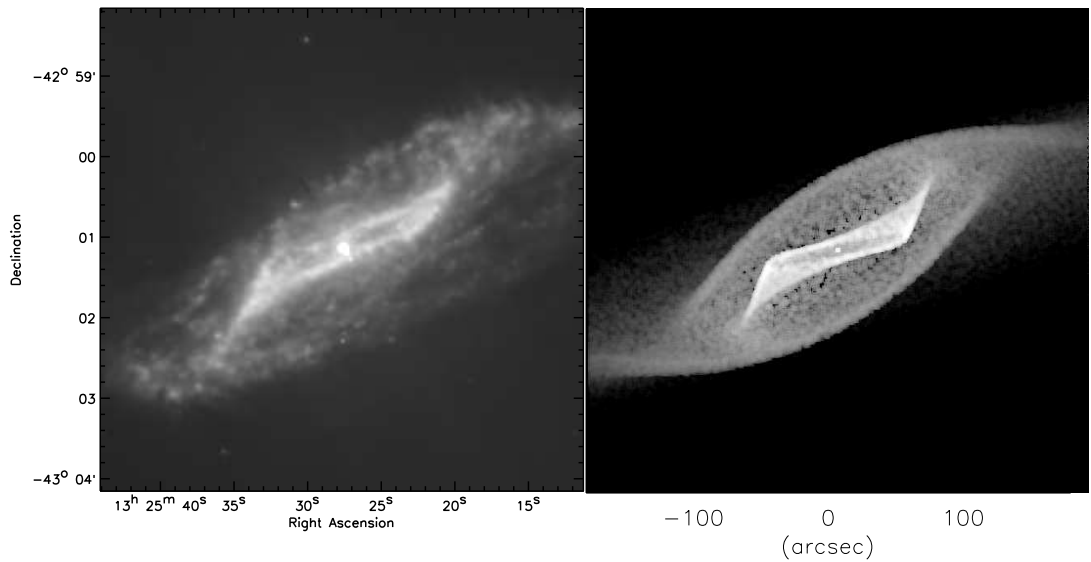}
\figcaption{
On the left is the {IRAC} $8.0\mu$m image.  On the right is the model 
for the warped disk with parameters given in Table 1. 
\label{fig:model}
}
\end{figure*}

\begin{figure*}
\plotone{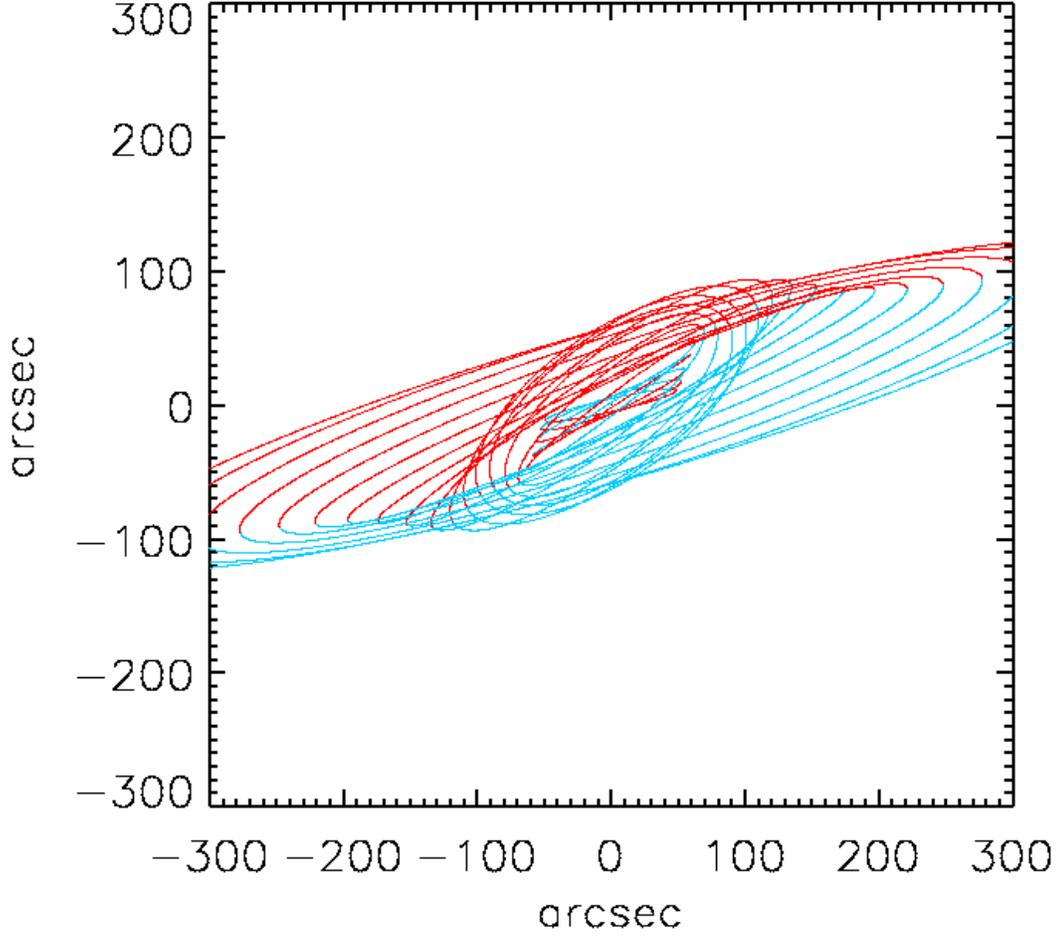}  
\figcaption{
This image shows rings for the warped disk model shown in the
previous figure.  The semi-circle of each ring nearest the observer is shown 
in red, that more distant shown in blue.
{\it For the black-white version of this figure:
The semi-circle of each ring nearest the observer is shown
in black, that more distant shown in gray.}
At near-infrared and visible wavelengths dust nearer the observer 
absorbs more background galactic starlight than dust more distant
from the observer. 
The inner fold corresponds to
the extinction feature south-east of the nucleus seen in
the J/H-band color map (see Fig.\ref{fig:colors}b).
The outer fold seen here corresponds to the extinction feature north
of the nucleus that is the northern edge of the dustlane
prominent in visible images of the galaxy. 
Regions where rings are close together correspond to regions of higher
surface brightness in the mid-infrared images.  Though the nearer
side can cause deeper extinction features in the near-infrared images,
both sides of the disk cause emission features in the mid-infrared images.
\label{fig:ring}
}
\end{figure*}

\begin{figure*}
\plotone{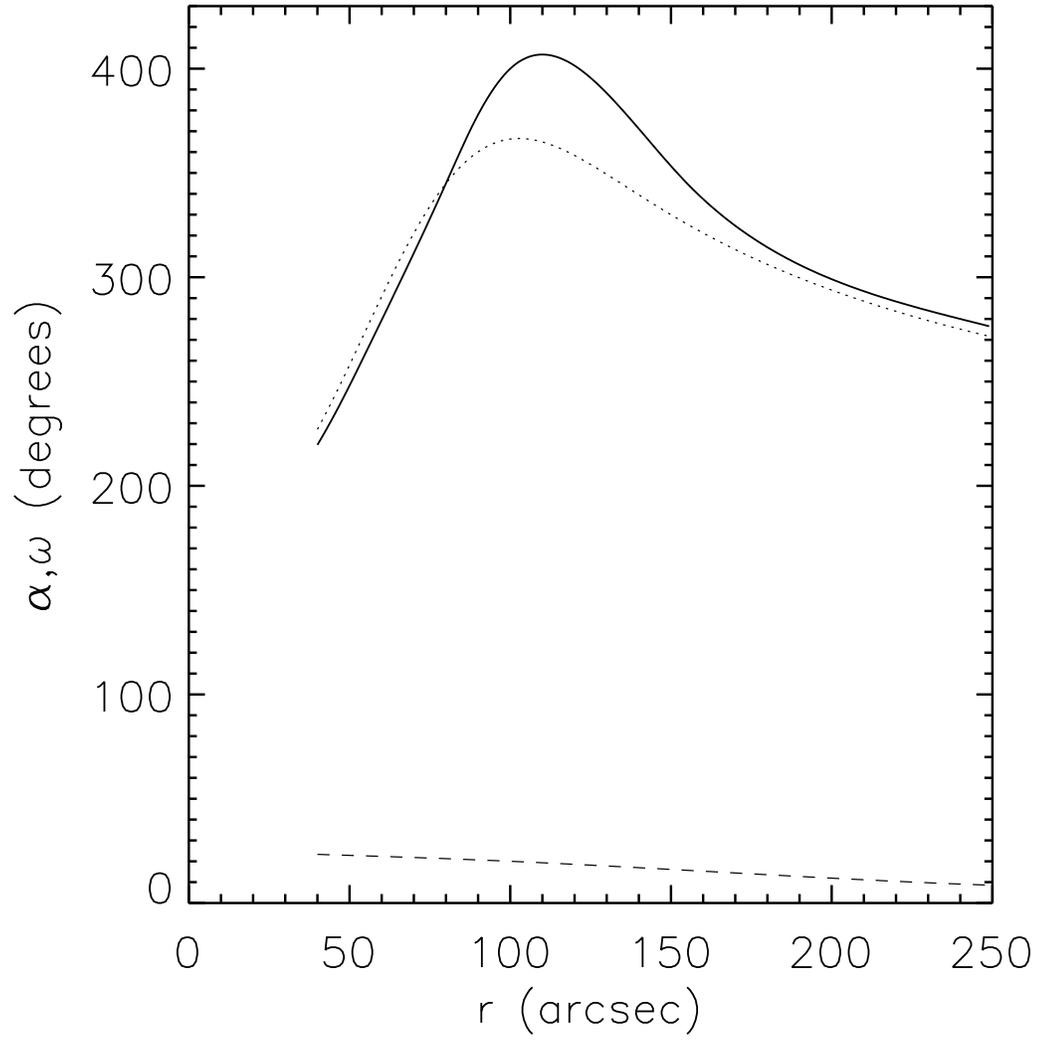}
\figcaption{
The solid line shows the precession angle, $\alpha(r)$ as a function of
radius used to make the model shown in Figure~\ref{fig:model}.
The dashed line shows the adopted inclination angle.
For comparison we show as a dotted line 
the precession angle used for the previous model
by \citet{quillen} (based on the near-infrared imaging).
These models should be approximately
consistent with the CO velocity field by \citet{quillenCO} 
and H\,$\alpha$ velocity field by \citet{nicholson92}.
\label{fig:alpha}
}
\end{figure*}

\begin{figure*}
\plottwo{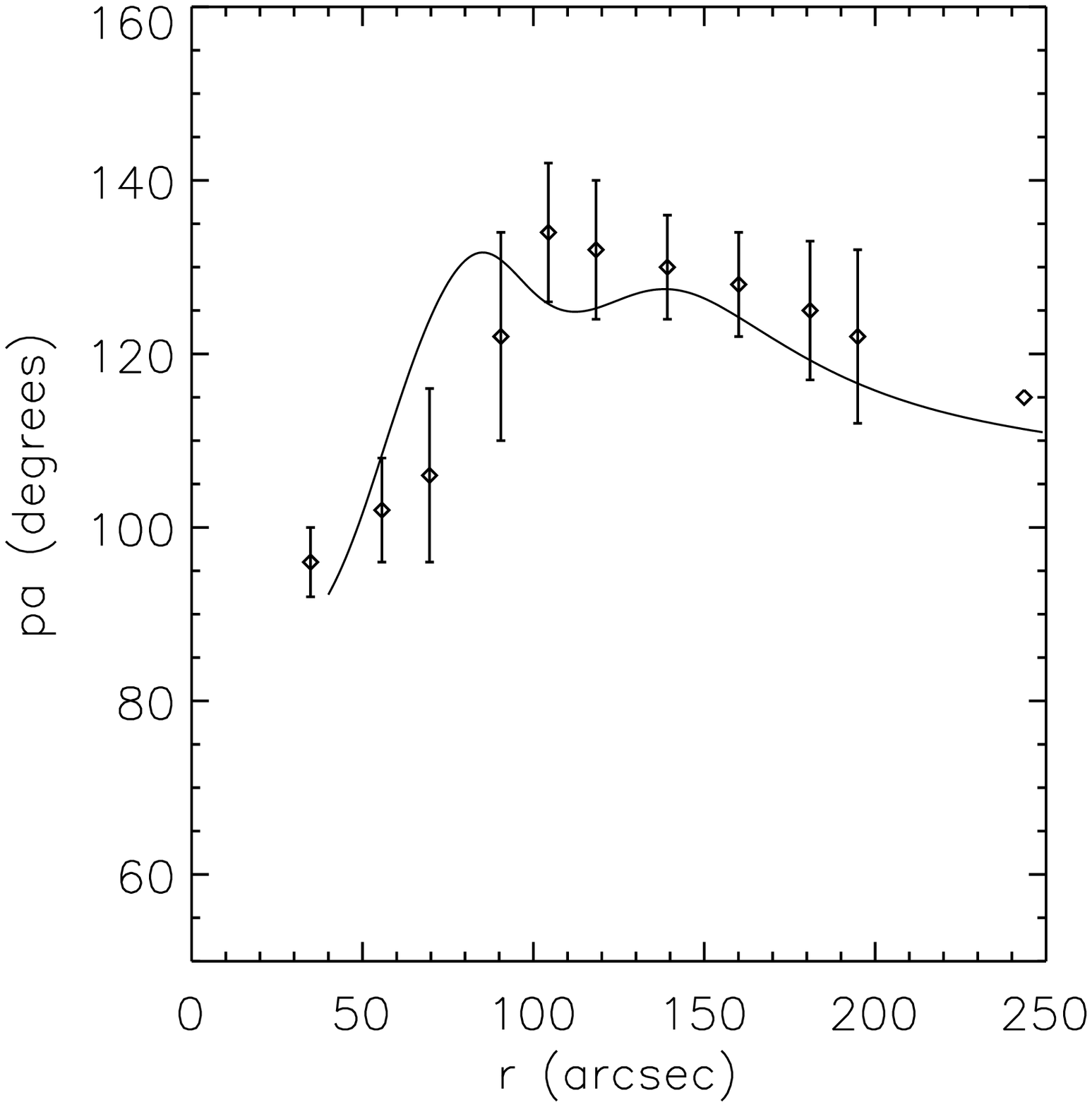}{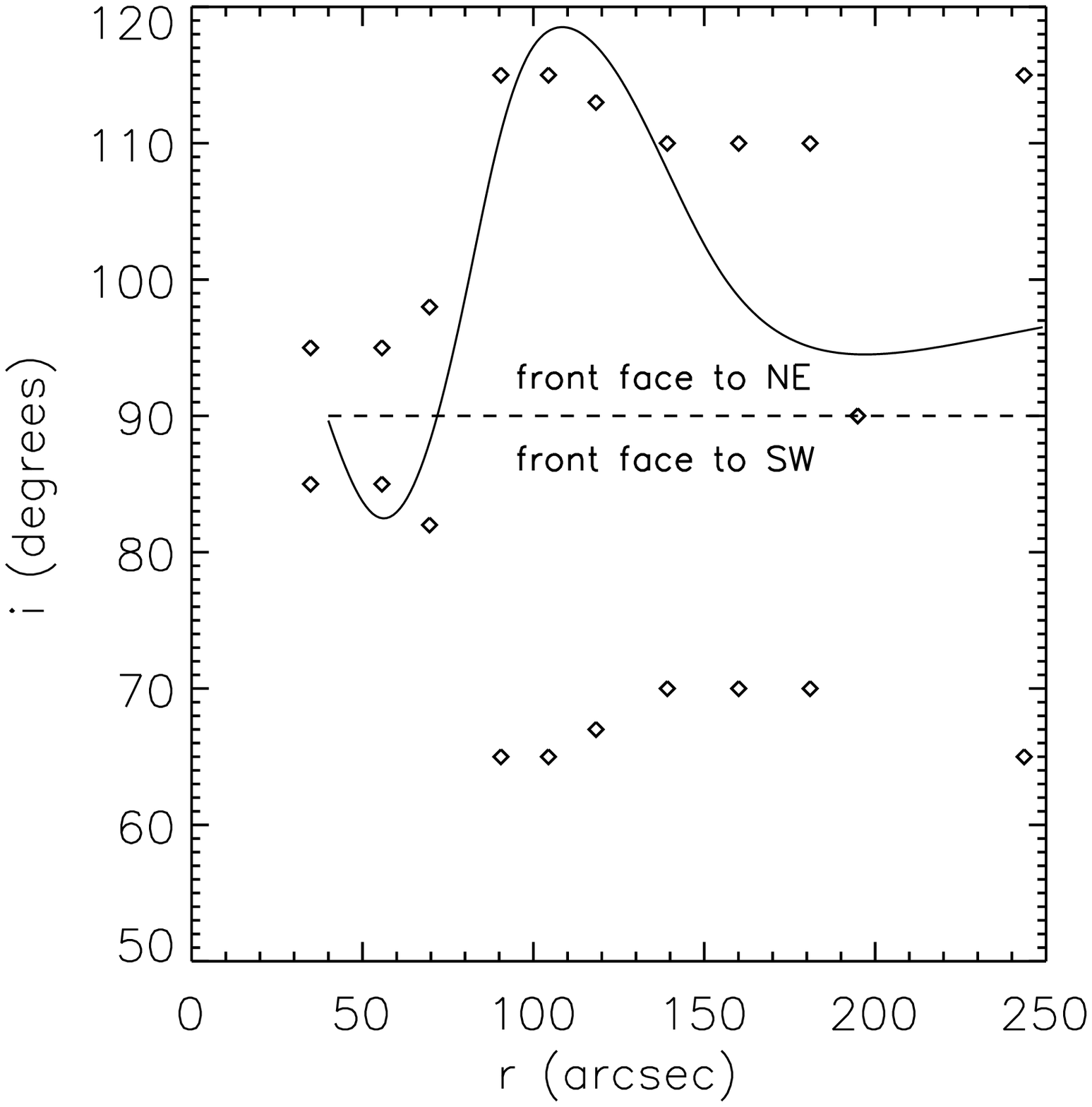}
\figcaption{
a) The position angle of each ring on the sky (anti-clockwise from 
North) is shown.
These angles are compared
with those estimated from tilted ring fits 
from the H\,$\alpha$ velocity field by \citet{nicholson92} 
that are shown as diamonds.  These points were taken from their Figure~9a.
b) We show the inclination angle of each ring 
with respect to the viewer for the
model shown in the previous figures.
The inclination of an edge-on disk is shown as a dotted
line at $90^\circ$.  Rings with inclinations above $90^\circ$ are oriented
with their nearest points to the north-east of the nucleus, and
those with inclinations below $90^\circ$ have nearest points 
to the south-west of the nucleus.
These angles are also compared
with those from the H\,$\alpha$ velocity field  (diamonds).
\citet{bland87,nicholson92} inferred that at $r\lesssim 60\arcs$ the front face
of the disk is oriented to the south-west of the nucleus, and vice versa
for $r\gtrsim 60\arcs$ (also see \citealt{sparke96}).
\label{fig:nichol}
}
\end{figure*}

\begin{figure*}
\plotone{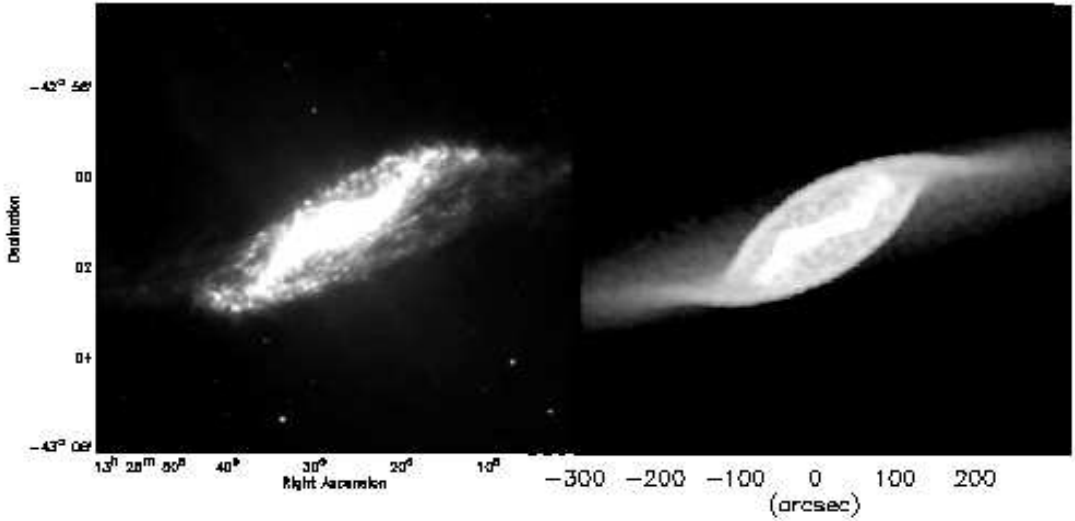}
\figcaption{
Similar to Figure~\ref{fig:model} except shown on a larger scale.
\label{fig:bigmodel}
}
\end{figure*}

\begin{figure*}
\plotone{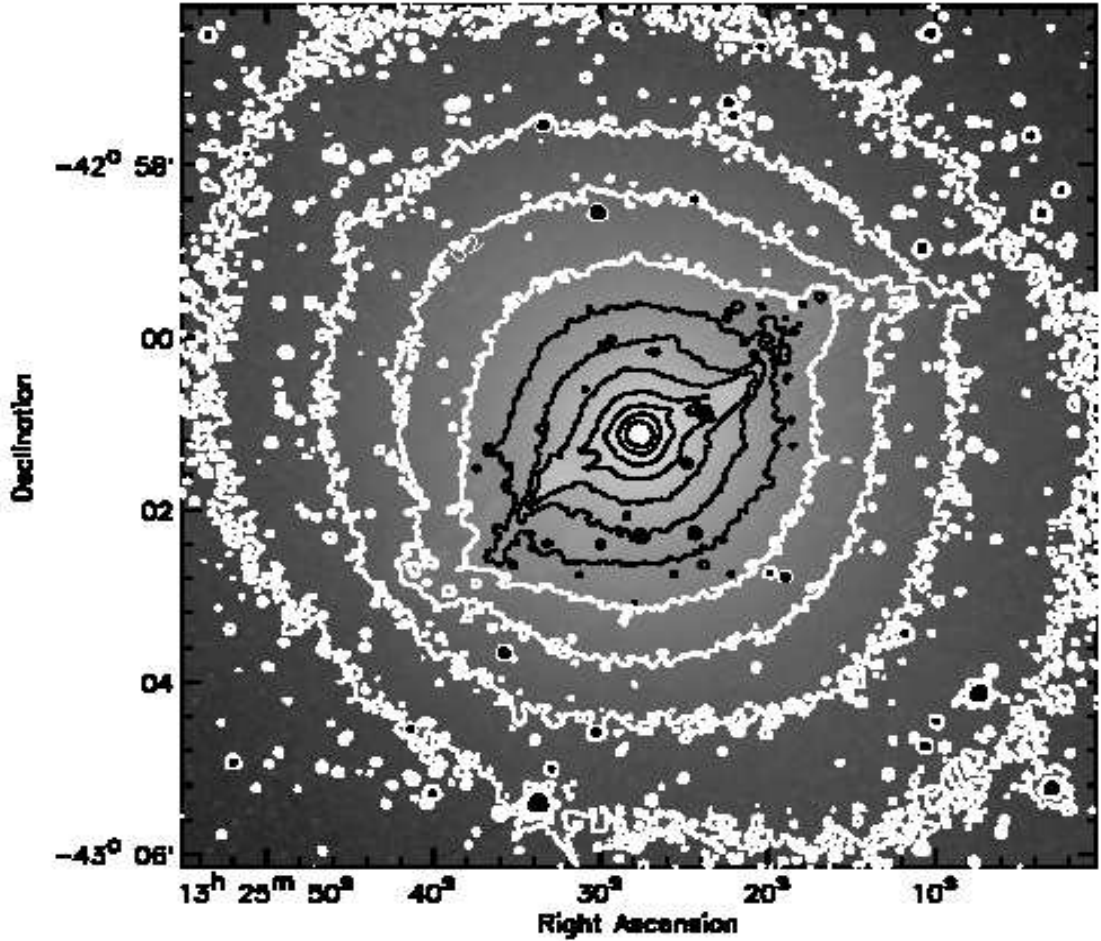}
\figcaption{
The {IRAC} Band 1 ($3.6 \mu$m) image shown with isophotal contours.
The lowest contour is at 0.63\,MJy\,sr$^{-1}$.  The 
contours are separated by 0.5 magnitudes or a factor of 1.585.
The galaxy is more highly elliptical at larger radii.
\label{fig:ch1}
}
\end{figure*}

\begin{figure*}
\plotone{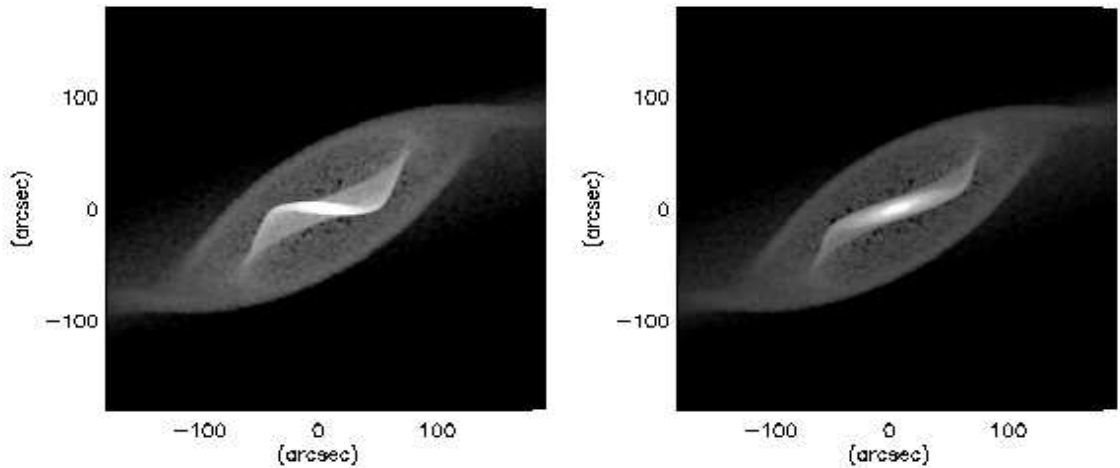}
\figcaption{
a) This model is similar to that shown in Figure~\ref{fig:model} except
there is no deficit in the dust distribution within $r_{gap}$.
b) This model is similar to that shown in Figure~\ref{fig:model} except 
the precession angle $\alpha(r)$ remains above $270^\circ$
for $r<50\arcs$.
The {IRAC} images are best matched with a model that contains
a deficit in the dust distribution within $r \sim 50\arcs$.
\label{fig:model_nohole}
}
\end{figure*}

The model that is most similar to the IRAC images (shown in
Fig.~\ref{fig:model}) contains a gap in the dust distribution with 
outer radius $r\sim 50\arcs$.  We compare the model shown in
Figure~\ref{fig:model} to a similar one that lacks the deficit in the dust
distribution.  This model is shown in Figure~\ref{fig:model_nohole}a,
and does not match the observed morphology as well as that containing a gap in
the dust distribution.   The smooth continuation of the precession angle into
the inner region 
results in an edge-on disk with respect to the viewer at some point within
$r=60\arcs$.  This causes the sharp bright linear feature at $r<60\arcs$ seen in
Figure~\ref{fig:model_nohole}a that is not exhibited by the {IRAC} images.

If the disk has a lower inclination with respect to the viewer, the observed
surface brightness is reduced.  We consider the possibility that the disk
precession angle differs in the inner region from that expected from a smooth
continuation of our model.  To test this possibility, we computed a model that
has a flattened (non-twisted) disk at $r \lesssim 60\arcs$.  This model is
shown in Figure~\ref{fig:model_nohole}b.  The sharp, bright, inner feature along
the east-west direction seen in Figure~\ref{fig:model_nohole}a is not evident
in Figure~\ref{fig:model_nohole}b.   However, the morphology of this model also
does not display the characteristics of our preferred model shown in
Figure~\ref{fig:model} that better resembles  the {IRAC} images.  In particular,
the triangular features to the south-east and north-west of the nucleus that
are seen in the near-infrared color maps on the south-east side \citep{quillen}
are not as good a match to those  of the IRAC images.   To remove the bright
inner feature of Figure~\ref{fig:model_nohole}a, the precession angle must
remain above $\alpha > 270^\circ$.  However, the curved triangle edge of
emission in the parallelogram is not present if the precession angle does not
steeply drop between a radius of $100\arcs$ and $50\arcs$.  We have tried models
with both increasing and decreasing precession angles near the nucleus, finding
no improved match to the observed morphology.  Our best match is the model
with an inner gap in the dust distribution.  We explored models with a lower but
constant surface density within the estimated gap radius of $\sim 50\arcs$, and
can exclude those with a surface density in the gap that is above 1/5 of
that at the outer gap edge.

Previous studies have discussed the possibility of a deficit in the gas
distribution in the same region as we find a deficit in the dust distribution.
At 100--200\,pc from the nucleus there is a circumnuclear molecular
disk that has been studied in molecular line emission \citep{israel90,israel91}.
A deficit in the ionized gas distribution outside the circumnuclear disk was
seen in the Pa\,$\alpha$ kinematics by \citet{marconi01}. In Pa\,$\alpha$, the
circumnuclear disk at a radius of $r \sim 6\arcs$ is seen, and so is emission at
significantly lower velocities (hence inferred larger radii) and significantly
higher velocities (within the sphere of influence of the massive black hole).
However Pa\,$\alpha$ emission is lacking at intermediate radii and velocities.
This lack of emission  would be expected if there were a deficit of gaseous
material between the radii of $\sim 6\arcs$ (set by the estimated outer radius
of the circumnuclear disk, \citealt{schreier96,marconi00,marconi01}) and
$50\arcs$ (estimated from our model). 

The rotation curve previously fit to the CO and H\,$\alpha$ kinematics rose
linearly (solid body) within a radius of 1\arcm\ of the nucleus
\citep{quillenCO,nicholson92}. Because an edge-on gas ring appears linear on a 
position-velocity diagram, a gas disk with an inner hole can mimic or be
confused with a gas disk extending all the way to the nucleus in a galaxy with
a linearly rising rotation curve.  \citet{nicholson92} showed that there was
a discrepancy between their measured rotation curve and that expected from a
$r^{1/4}$ or deVaucoulers law.  They listed a possible hole in the HII region
distribution as a possible cause for this discrepancy.  A linearly rising
rotation curve within a radius of 1\arcm\ of the nucleus is not consistent 
with the K-band surface brightness profile (see the rotation curve predicted in
Figure~13 by \citealt{marconi06}).  The apparent region of solid body rotation
need not be accounted for with a galactic bar, as proposed by \citet{mirabel99},
and can be better explained by a gap in the gas and dust distribution.


\section{Summary and Discussion}

By integrating the light through an emitting, optically thin, dusty, and warped
disk, we have successfully matched the morphology of Centaurus\,A seen in
mid-infrared {IRAC} images.  We confirm previous proposals that the disk
morphology is well explained by a warped disk
\citep{bland86,bland87,quillen,nicholson92,leeuw02} rather than a barred one
\citep{mirabel99}.  The disk is nearly edge-on with respect to the viewer, but
tilts so that folds appear above and below the galaxy equator.  There is a fold
south of the nucleus at a radius $r\sim 60\arcs$ responsible for high
extinction seen in near-infrared images, and a fold north of the nucleus at
larger radii responsible for the northern edge of the dust lane seen in optical
and images.  Extinction features seen in the near-infrared extinction map
correspond to folds in the disk that are located nearer the observer and so
absorb more background stellar light from the galaxy.  In the mid
infrared, however, folds on both the near and opposite side of the galaxy
correspond to bright emission features.
The inner folds account for the parallelogram shape, while the outer folds
account for the northern edge of the dust lane seen in optical images and a
fainter oval of emission seen outside the parallelogram in the IRAC
images.   

The disk geometry we use to match the mid-infrared morphology is similar to that
found previously by \citet{quillen},  and is also similar to that required
to fit the CO and H\,$\alpha$ velocity field
\citep{bland87,quillenCO,nicholson92}. The geometric warp models by
\citet{quillen,nicholson92} have been predictive; they provide good matches to
the mid-infrared morphology.  Some differences in the precession angle exist in
the model at $r \gtrsim 100\arcs$ compared to previous work.
Previous CO, H\,$\alpha$ spectra and near-infrared imaging were not
sensitive enough to provide tight kinematic constraints on the outer disk.
The {\it Spitzer\/} data have allowed us to extend the model past $r\sim
100\arcs$ compared to previous models and see closer in to the nucleus 
where the extinction is high at shorter wavelengths.  Better constraints on the
disk geometry could be achieved in the 
future by fitting observations at more than
one wavelength.  For example, modern high resolution kinematic observations that
are fit along with the {\it Spitzer\/} data using the same model would allow
much stronger constraints on the disk geometry and gas and dust distribution
than the slight modification to previous geometric models that we 
have used here.

The warp disk model suggests that there is a gap in the dusty
disk at $6\arcs \lesssim r\lesssim 50\arcs$.   
A gap exists in same region in the gas distribution.  \citet{marconi01} saw
a deficit of Pa\,$\alpha$ emission near the nucleus at intermediate
velocities.   \citet{nicholson92} suggested that the discrepancy between the
measured linearly rising rotation curve within a radius of $60\arcs$ might be
explained by a hole in the ionized gas distribution.  
It is not easy to determine
if there is a gap in the dust distribution since the infrared surface
brightness depends on the inclination of the disk and the disk is highly
corrugated.   If the disk twists to lower inclinations (closer to face-on) at
small radii, it would have a lower surface brightness near the nucleus.  We
have explored models with smoothly varying radial surface brightness
distributions, and smoothly varying precession and inclination angles.  We find
that only models with a deficit of dust interior to $\sim 50\arcs$ resemble the
mid-infrared images.  We exclude models that have more than $\sim 1/5$ the
surface density within $r=50\arcs$ as at that radius.  We conclude that there is
a gap in the gas and dust distribution between 0.1 and 0.8\,kpc from the
nucleus.  The inner radius of the gap we have taken from studies of the
circumnuclear disk \citep{marconi01,israel91}, and the outer radius is
estimated from our model.

It is interesting that only the region between the circumnuclear disk
(100--200\,pc) and $\sim 0.8$\,kpc has been depleted of gas and dust.  While we
find no evidence for a large scale stellar bar in Centaurus\,A, a dense
gas disk could have exhibited dynamical instabilities depositing gas into the
circumnuclear disk (e.g., \citealt{shlosman}).  Energetic star formation or
activity associated with the black hole can deplete and evacuate the central
region of a galaxy (e.g., \citealt{springel05,veilleux05}), though it is not
clear how a circumnuclear disk would be protected from or reformed following
this activity.  Nearby galaxies exhibit evacuated central regions or
circumnuclear rings.  For example, the Circinus galaxy has a $\sim 500$\,pc
radius molecular ring \citep{curran98} and contains a Seyfert nucleus.  M82
contains a 1\,kpc radius circumnuclear molecular ring;  however, a previous
epoch of star formation has occurred within this ring (e.g.,
\citealt{forster03}). 
Future observational studies may differentiate between the role of star
formation, dynamical instabilities and nuclear activity in disrupting the gas
and dusty disk in Centaurus\,A.

The best matching geometric warp model requires a change in the slope of the
precession angle at a radius of about $r\sim 100\arcs$.  Two previous models
account for this twist.  \citet{quillen} used a model in which the galaxy was
prolate in its outer region and the galaxy ellipticity abruptly dropped to 
zero within $r \sim 80\arcs$.  However, this abrupt drop is not consistent with
the isophote shapes at $3.6\,\mu$m.  A prolate model (with the advantage of a
relatively short timescale) might account for the disk geometry if modified to
include the self-gravity of the disk.  The polar ring model by \citet{sparke96}
naturally accounts for the reversal, but requires a longer timescale to
operate.  These dynamical models could be updated to use a more accurate mass
distribution and rotation curve. 
Improvements to these models may lead to better understanding of the galaxy
merger that created Centaurus\,A's peculiar morphology, as well as the merger's
role in feeding the active galactic nucleus. 

\acknowledgments

We thank Joss Bland-Hawthorn, Varoujan Gorjian 
and Vassilis Charmandaris for helpful comments and
suggestions.  We thank the Research School of Astronomy and Astrophysics of the
Australian National University and Mount Stromlo Observatory for hospitality and
support for ACQ during Spring 2005.  Support for this work was in part
provided by National Science Foundation grant AST-0406823, and the National
Aeronautics and Space Administration under Grant NNG04GM12G issued through
the Origins of Solar Systems Program.  We acknowledge support by award
HST-GO-10173-09.A through the Space Telescope Science Institute.

\clearpage


\smallskip

\begin{deluxetable}{lcc}
\tablewidth{0pt}
\tablecaption{Model parameters \label{mod_params}}
\tablehead{
\colhead{Description} &
\colhead{Parameter} &
\colhead{Value}
}
\startdata
\cutinhead{Galaxy principal axis}
$PA$ on sky           & $\chi$      & $20^\circ$  \\
Tilt                  & $\vartheta$ & $75^\circ$ \\
\cutinhead{Dust intensity}
Dust emissivity index & $\beta_e$   &  4.0   \\
Aspect ratio          & $k_{50} $   &  0.1   \\
Aspect ratio index    & $\beta_k$   &  0.9   \\
Radius of inner hole  & $r_{gap}$  &  $50\arcs$  \\
\enddata
\tablecomments{
The precession angle for this model, $\alpha(r)$, 
is a spline function that interpolates between
10 points in the format (radius in arcseconds, $\alpha$ in degrees):
(0.2, 180), (36.0, 210 ), (57, 270), (80, 345),
(89, 375), (100, 400), (135, 380), (155,345), (240,280), (400,240).
The inclination angle $\omega(r)$ is a spline function that
interpolates between the following points
(0.2, 25), (100, 20), (260, 8), (400, 8).}
\end{deluxetable}

\end{document}